\documentclass[twocolumn,usenatbib,usegraphicx ]{mn2e}

\def\lcdm{$\Lambda {\rm CDM} \,\,$} 
\def\mnras{MNRAS}
\def\prd{Physical Review D}
\def\aj{AJ}
\def\apj{ApJ}
\def\apjl{ApJL}
\def\apjs{ApJS}
\def\aap{A\&A}

\usepackage{epsfig}

\begin{document}
\title{The luminosity bias relation from filaments in the Sloan Digital Sky Survey Data Release Four}   
\author[B. Pandey and S. Bharadwaj] {Biswajit Pandey\thanks{Email: 
    pandey@cts.iitkgp.ernet.in} and Somnath Bharadwaj\thanks{Email: 
    somnathb@iitkgp.ac.in}
\\ Department of Physics and Meteorology\\ and
\\ Centre for Theoretical Studies \\ IIT Kharagpur \\ Pin: 721 302 ,
India }
\maketitle

\begin{abstract}
 We compare quantitative estimates of the filamentarity 
of the galaxy  distribution  in seven  nearly two dimensional sections  
from the survey against the predictions  of \lcdm N-body
simulations. The filamentarity of  the actual galaxy distribution is
known to be  luminosity dependent. It is also known that 
the filamentarity of the simulated galaxy distribution is  
highly sensitive to the bias, and the simulations are consistent with
the data for only a narrow range of bias.  We apply this
feature to  several volume limited subsamples with different
luminosities to determine a luminosity bias relation. The
relative bias $b/b^*$ as a  function  of the luminosity ratio $L/L^*$
is found to be well described by a straight line $b/b^*=A + B \,
(L/L^*)$ with $A=0.833 \pm 0.009$ and $B=0.171 \pm 0.009$. 
Comparing with  the earlier works  all of which use ratios of
the two-point statistics, we find that our results are 
consistent with \citet{nor} and \citet{teg2},  while a steeper
luminosity  dependence found by \citet{beno} is inconsistent. 
\end{abstract}
\begin{keywords}
methods: numerical - galaxies: statistics - 
cosmology: theory - cosmology: large scale structure of universe 
\end{keywords}

\section{Introduction}
Filaments are possibly the most prominent visible feature in galaxy
redshift surveys. The Las Campanas Redshift Survey (LCRS;
\citealt{shect}) has been analyzed showing the observed filamentarity
to be statistically significant on length-scales upto $\sim 80 h^{-1}
{\rm Mpc}$ and not beyond \citep{bharad2}, establishing the filaments
to be the largest known statistically significant coherent structures
in the galaxy distribution.  The Sloan Digital Sky Survey \citep{york}
is currently the largest galaxy redshift survey. Studies using two
nearly two-dimensional (2D) sections from the Sloan Digital Sky Survey
Data Release one (SDSS DR1; \citealt{abaz1}) confirm the earlier result
\citep{pandey}.  This analysis also shows the filamentarity to be
luminosity dependent with the brighter galaxies having a more compact
and less filamentary distribution as compared to the faint ones.  A
more recent study \citep{pandey1} using a larger data-set from the
Sloan Digital Sky Survey Data Release Four (SDSS DR4; \citealt{adel})
confirms the luminosity dependence and shows this effect to be
considerably enhanced for galaxies brighter than $L^*$.  This study
also shows the filamentarity to be dependent on the colour and
morphology of the galaxies.

The possibility that the galaxies are a biased tracer of the
underlying dark matter distribution is now quite widely accepted. A
study comparing the filamentarity observed in the LCRS with the
prediction of \lcdm dark matter N-body simulations finds the N-body
predictions to be very sensitive to the bias parameter
\citep{bharad3}. Increasing the bias causes the simulated galaxies
to get concentrated into smaller regions which correspond to the peaks
in the dark matter distribution.  The large-scale filamentarity of the
simulated galaxy distribution is found to decrease with increasing
bias. The filamentarity of the simulated galaxy distribution and the
LCRS galaxies in the absolute magnitudes range $-21.5 \le M \le -20.5$
are found to be consistent for a bias $b=1.15$.  An unbiased \lcdm
model $(b=1)$ and a large bias parameter $(b=1.5)$ are both ruled out.
Comparing the filamentarity in biased N-body simulations with the
actual galaxy distribution provides a novel technique to determine the
galaxy bias.

The large coverage and high galaxy number density of the SDSS permits
us to study the filamentarity as a function of galaxy luminosity. In
this {\em Letter} we compare the filamentarity of galaxy samples in
different luminosity bins drawn from the SDSS DR4 against the
predictions of biased N-body simulations and use this to determine the
bias as a function of luminosity {\em ie.} a luminosity bias relation.

The luminosity dependence of galaxy clustering is an important issue
with considerable implications for theories of galaxy formation.
Observations show that the luminous galaxies exhibit a stronger
clustering than their fainter counterparts (e.g. \citealt{hamil},
\citealt{davis1}, \citealt{white}, \citealt{park}, \citealt{love},
\citealt{guzo}, \citealt{beno}, \citealt{nor}, \citealt{zehavi}). It
has remained difficult to quantify this effect through a luminosity
bias relation because of the limited dynamical range of even the
largest redshift surveys. Traditionally the luminosity bias relation
has been determined by comparing observations of the two-point
statistics, namely the correlation function or the power spectrum in
different luminosity bins. This allows the relative bias $b/b^*$
to be studied as a function of the luminosity ratio $L/L^*$ 
 where $b^*$ is the bias corresponding to the galaxies with
the characteristic luminosity $L^*$ or equivalently the characteristic
magnitude $M^*$. 
  To our knowledge there are three earlier results.
\citet{beno} have analyzed the luminosity dependence of the bias
in the  SSRS2 \citep{dacosta}  which has been found to be well fitted
by a luminosity bias  relation $(b/b^*) = 0.7 + 0.3 \, (L/L^*)$  
\citep{pkok}. The analysis of the 2dFGRS \citep{colless} by 
\citet{nor} yields a more modest luminosity dependence $(b/b^*) =
0.85+ 0.15 (L/L^*)$. \citet{teg2} have analyzed the SDSS \citep{abaz2}
to obtain a relation $(b/b^*) = 0.895+ 0.15
(L/L^*)-0.04(M-M^*)$. Unlike these estimates, the present analysis
uses a global property namely the filamentarity, to determine the
luminosity bias relation.

There currently exist several ways to directly determine the bias.
Measuring the redshift space distortion parameter
$\beta=\Omega_m^{0.6}/b$ (e.g. 2dFGRS, \citealt{haw}; SDSS,
\citealt{teg2}) in combination with an independent determination of
$\Omega_{m0}$ allows $b$ to be determined. The bispectrum (2dFGRS,
\citealt{verde}) provides a technique to determine the bias from
redshift surveys without the need of inputs from other observations. A
combination of weak lensing and the SDSS galaxy survey has been used
by \citet{sel} to determine the bias. We note that most of these
methods require large galaxy samples with a high number density.  It
has not been possible to apply these techniques to subsamples with
different luminosities and determine a luminosity bias relation.

\section{Data and method of analysis}

\begin{figure}
\rotatebox{-90}{\scalebox{.4}{\includegraphics{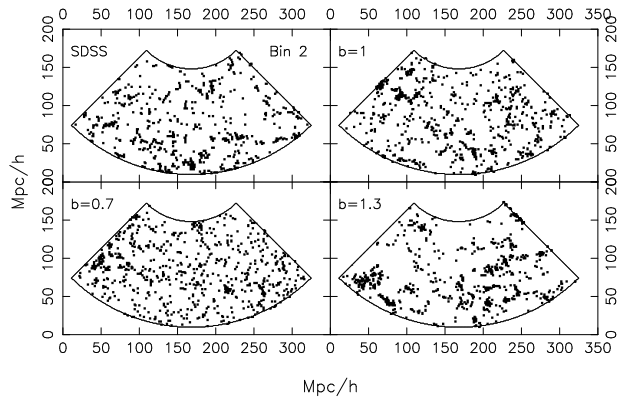}}}
\caption{This shows the galaxy distribution in one of the SDSS strip
  in Bin2 along with some of the simulated slices with bias values as
  indicated in the panels.   
}
\label{fig:1}
\end{figure}

 For the present analysis we use seven non-overlapping strips each with
 sky coverage $2^{\circ} \times 90^{\circ}$ which lie entirely within
 the survey area of SDSS DR4. These strips are identical in sky
 coverage as the ones used in \citet{pandey1} and are shown in Figure
 1 of that paper. For each strip we have extracted five different
 volume limited subsamples, 'Bin 1' ... 'Bin 5', applying the absolute
 magnitude and redshift limits listed in Table~1. The thickness of the
 resulting subsamples increases with redshift. For our analysis we
 have considered a smaller region of uniform thickness corresponding
 to the value at the lowest redshift. For each bin the area and
 thickness are listed in Table~1. For all the subsamples, the
 thickness is much smaller than the other two dimensions and hence it
 is collapsed along the thickness resulting in a 2D distribution
 (Figure \ref{fig:1}). For each luminosity bin the galaxy number
 density varies slightly across the seven strips and the average
 density along with the $1-\sigma$ variation is shown in Table~1. It
 may be noted that the bins span $\Delta M= 1$ in absolute magnitude
 and are in order of increasing luminosity. The characteristic
 magnitude $M^*$ has a value $M^*=-20.44\pm0.01$ for the SDSS
 \citep{blan3} and the $M^*$ galaxies appear in Bin 3 of our
 sub-samples. The galaxy number density falls considerably in the bins
 containing galaxies brighter than $M^*$.

\begin{table*}{}
\caption{This shows the absolute magnitude and redshift limits for the
  different volume limited subsamples analyzed. The median magnitude,
  area, thickness and average galaxy number density with $1-\sigma$
  variations from the $7$ strips are also shown.}\vspace*{.2cm}
\begin{tabular}{|c|c|c|c|c|c|}
\hline Bin & Magnitude range & Redshift range &
 Area & Thickness & Density\\ 
&&&$10^{4}\, h^{-2}\, {\rm Mpc^2}$&
 $h^{-1} \, {\rm Mpc}$ & $10^{-2}\, h^{2} {\rm Mpc}^{-2}$\\ \hline 
Bin 1 &$-18.5 \geq M_r \geq -19.5$ & $0.022\leq z \leq 0.061$ & $2.21$ & $2.33$ & $1.93 \pm 0.16$ \\ 
Bin 2 &$-19 \geq M_r \geq -20$ & $0.028\leq z \leq 0.075$ &  $3.28$ & $2.91$ & $2.33 \pm
 0.2$ \\ 
Bin 3 &$-20\geq M_r \geq -21$ & $0.043 \leq z \leq 0.114$ & 
 $7.41$ & $4.52$ & $1.77 \pm 0.15$ \\ 
Bin 4 &$-21 \geq M_r \geq -22$ & 
 $0.067\leq z \leq 0.169$ & $15.5$ & $6.94$ & $ 0.67 \pm 0.04 $ \\
Bin 5 &$-21.5 \geq M_r \geq -22.5$ & $0.083\leq z \leq 0.2$ & 
 $20.9$ & $8.55$ & $ 0.25 \pm 0.01$ \\ \hline
\end{tabular}
\end{table*}

We simulate the dark matter distribution using a Particle-Mesh (PM)
N-body code. The simulations use $256^3$ particles on a $512^3$ mesh,
and they have a comoving volume $[921.6 h^{-1} {\rm Mpc}]^3$ . We use
$(\Omega_{m0},\Omega_{\Lambda0},h)=(0.3,0.7,0.7)$ for the cosmological
parameters along with a \lcdm power spectrum with spectral index
$n_s=1$ and normalization $\sigma_8=0.84$ (\citealt{sperg}). 
 A ``sharp cutoff'' biasing,  scheme
\citep{cole} was used to extract particles  from the N-body 
simulations. Identifying these particles as galaxies, we  have 
galaxy distributions that  are 
biased relative to the dark matter. The bias parameter
$b$ of each simulated galaxy sample was estimated using the ratio
\begin{equation}
b=\sqrt{ \frac {\xi_g(r)}{\xi(r)}}
\end{equation}
where $\xi_g(r)$ and $\xi(r)$ are the galaxy and dark matter two-point
correlation functions respectively. This ratio is found to be constant
at length-scales $r \ge 5 h^{-1} {\rm Mpc}$ and we use the average
value over $5-40 h^{-1}{\rm Mpc}$. We use this method to generate
galaxy samples with bias values in the range 1 to 2 in steps of
0.1. Galaxy distributions with bias $0 < b < 1$ were generated by
adding randomly distributed particles to the dark matter
distribution. In addition to the bias, the filamentarity of the
simulated galaxy distribution is also expected to depend on the degree
of non-linearity in the evolution (see \citealt{verde} for effects on
the bispectrum).  To asess this effect we have run simulations using
$\sigma_8=1$ and the best fit cosmological parameters determined from
the SDSS+WMAP 3 year data (\citealt{sperg1}, $\sigma_8=0.77$), and
find that variations in this range do not cause a statistically
significant change in the filamentarity.

 The peculiar velocity effects were included to produce galaxy
distributions in redshift space. We have used three independent
realisations of the N-body simulations, and for each luminosity bin we
have extracted six different simulated data-sets from each
simulation. This gives us a total of eighteen simulated galaxy
distribution for each luminosity bin. The simulated galaxy
distributions (Figure 
\ref{fig:1}) have the same area, thickness and the mean number density
as those of the corresponding bins (Table~1). The simulated data were
analyzed in exactly the same way as the actual data.

All the sub-samples that we have analyzed are nearly two dimensional.
The samples were all collapsed along the thickness (the smallest
dimension) to produce 2D galaxy distributions (Figure \ref{fig:1}). We
use the 2D ``Shapefinder'' statistic \citep{bharad1} to quantify the
average filamentarity of the patterns in the resulting galaxy
distribution.  A detailed discussion of the method of analysis is
presented in \citet{pandey1} (and references therein), and we
highlight only the salient features here. The reader is referred to
\citet{sahni} for a discussion of Shapefinders in three
dimensions.

 The galaxy distribution is represented as a set of 1s on a 2-D
 rectangular grid of spacing $1 \, h^{-1} {\rm Mpc} \times 1 \,h^{-1}
 {\rm Mpc}$. The empty cells are assigned a value $0$. We identify
 connected cells having a value 1 as clusters using the
 'Friends-of-Friend' (FOF) algorithm. The filamentarity of each
 cluster is quantified using the Shapefinder ${\cal F}$ defined as
\begin{equation}
{\cal F} = \frac{(P^2 - 16 S)}{(P-4 l)^2}
\end{equation}
 where $P$ and $S$ are respectively the perimeter and the area of the
 cluster, and $l$ is the grid spacing. The  Shapefinder ${\cal F}$
 has  values $0$ and $1$ for a square and filament respectively, and
 it assumes intermediate values as  a  square is deformed to a
 filament. We use the average filamentarity  
\begin{equation} 
F_2 = {\sum_{i} {\cal S}_i^2 {\cal F}_i\over\sum_{i}{\cal S}_i^2} \,. 
\end{equation}
to asses the overall filamentarity of  the clusters in the galaxy
distribution. 

\begin{figure}
\rotatebox{-90}{\scalebox{.35}{\includegraphics{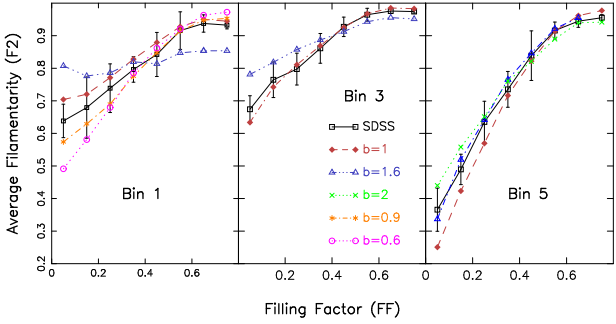}}}
\caption{ This shows the Average Filamentarity ($F_2$) as a function
 of Filling Factor ($FF$) for every alternate luminosity bins
 (Table~1). The actual data along with the corresponding results from
 \lcdm N-body simulations are shown for a few representative values of
 the bias. We use only the range $0.05\leq FF \leq 0.65$ in our
 analysis.}
\label{fig:2}
\end{figure}

A problem arises because the distribution of 1s corresponding to the
galaxies is rather sparse. Only $\sim 1 \%$ of the cells contain
galaxies and there are very few filled cells which are interconnected.
As a consequence FOF fails to identify the large coherent structures
which are visually discernible as filaments in the galaxy distribution
(Figure \ref{fig:1}). We overcome this problem by successively
coarse-graining the galaxy distribution. In each iteration of
coarse-graining all the empty cells adjacent to a filled cell are
assigned a value $1$. This causes clusters to grow, first because of
the growth of individual filled cells, and then by the merger of
adjacent clusters as they overlap. Coherent structures extending
across progressively larger length-scales are identified in
consecutive iterations of coarse-graining.  So as not to restrict our
analysis to an arbitrarily chosen level of coarse-graining, we study
the average filamentarity after each iteration of coarse-graining. The
filling factor $FF$ quantifies the fraction of cells that are filled
and its value increases from $\sim 0.01$ and approaches $1$ as the
coarse-graining proceeds.  We study the average filamentarity $F_2$ as
a function of the filling factor $FF$ (Figure \ref{fig:2}) as a
quantitative measure of the filamentarity at different levels of
coarse-graining. The values of $FF$ corresponding to a particular
level of coarse-graining shows a slight variation across the
data-sets. In order to combine and compare the results from different
data-sets, for each data-set we have interpolated  $F_2$ to $7$
values of $FF$ at an uniform spacing of  $0.1$ over the interval
$0.05$ to $0.65$. Further coarse-graining beyond $FF \sim 0.65$
washes away the filaments and hence we do not include this range for 
our analysis. 

For each luminosity bin (Table~1) the seven different strips were used
to determine the mean and the variance-covariance matrix for $F_2$ as
a function of $FF$. This was compared against the mean and the
variance-covariance matrix estimated from eighteen simulated data-sets
for each value of the bias. Bias values in the range $0.5 - 2.0$ were
considered.  For each value of the bias we use the Hotelling's two
sample $T^2$ test \citep{krzn} to test the null hypothesis that the
actual values of $F_2$ as a function of $FF$, and the predictions of
the simulated data belong to the same statistical sample.  The $T^2$
statistic is defined as
\begin{equation}
T^2=(\bar{X_1}-\bar{X_2})^{\prime}\
\Big[\Big(\frac{1}{N_1}+\frac{1}{N_2}\Big)S_p\Big]^{-1}\ (\bar{X_1}-\bar{X_2}) 
\label{eq:2}
\end{equation}
where $\bar{X_1}$ and $\bar{X_2}$ are the vectors containing the
sample means and $N_1=7$ and $N_2=18$ are the number of realizations for
the data and the simulation respectively. $S_p$ is the pooled covariance
matrix defined as $S_p= \frac {n_{1} S_{1}+n_{2} S_{2}}
{n_{1}+n_{2}}$ where $n_{1}=N_{1}-1$, $n_{2}=N_{2}-1$ and $S_{1}$, and
$S_{2}$ are the variance-covariance matrix for the data and model
respectively. 
 The null hypothesis is rejected for an observed value
$T_{0}^2 > \frac {(n_{1}+n_{2})p} {(n_{1}+n_{2}-p+1)} \,
F(p, n_{1}+n_{2}-p+1, 1-\alpha)$ where $p=7$ is the number of variables
and $F$ is the $F$ distribution.  We choose $\alpha=0.05$ which allows
us to reject  bias 
values at the $95 \%$ confidence level. For each luminosity bin, bias
values for which  the observed $T_0^2$ is greater than $T^2=24.758$
are rejected with  $95 \%$ confidence. 

We have tested for possible instabilities in inverting the pooled 
covariance matrix  $S_p$ arising from   the limited  number of  realisations
 used  to estimate $S_1$  and $S_2$ \citep{hart} 
 by repeating the analysis using a smaller number of realisations
 ($N_1=4$ and $N_2=10$). The results are found to be 
unchanged, except for a somewhat larger  $b$ range that is accepted
with $95 \%$  confidence. 

\begin{figure}
\rotatebox{-90}{\scalebox{.32}{\includegraphics{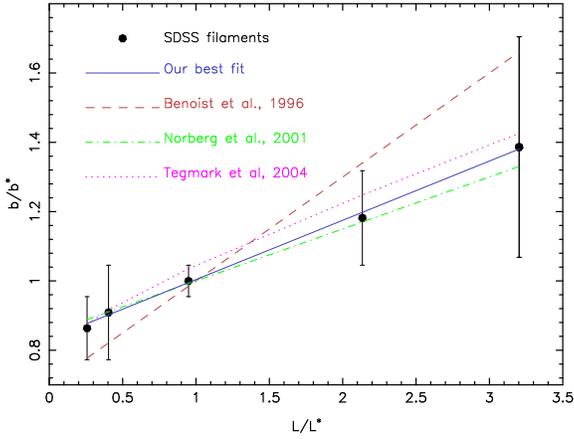}}}
\caption{ This shows the measured relative bias $\frac{b}{b^*}$ of the
SDSS galaxies as a function of $\frac{L}{L^*}$. The solid curve shows
the best fit curve to our data $\frac {b}{b^*} = 0.833 + 0.171
\frac{L}{L^*}$. Fits to data from earlier works are also shown for
comparison. The error-bars shown in the figure correspond to $95\%$
confidence intervals.}
\label{fig:3}
\end{figure}

\begin{table}{}
\caption{For each luminosity bin this shows the median magnitude and
the corresponding bias determined from our analysis. The errors in the
bias indicate the $95\%$ confidence interval.}
\vspace*{.2cm}
\begin{tabular}{|c|c|c|}
\hline 
Bin &Median magnitude&Bias\\
\hline
 Bin 1&$-18.96$&$0.95\pm0.1$ \\
 Bin 2&$-19.45$&$1.0\pm0.15$ \\ 
 Bin 3&$-20.38$&$1.1\pm0.05$ \\ 
 Bin 4&$-21.26$&$1.3\pm0.15$ \\
 Bin 5&$-21.7$&$1.52\pm0.35$ \\  
 \hline
\end{tabular}
\end{table}

\section{Results and Conclusions}
 We show results for the Average Filamentarity $F_2$ as a function of
 the Filling Factor $FF$ in Figure \ref{fig:2}. Increasing the
 coarse-graining causes adjacent clusters to connect up into
 progressively longer interconnected filaments and as a consequence
 $F_2$ increases with $FF$ in the simulations as well as the actual
 data. At $FF$ in the range $0.5 - 0.6$ there is a percolation
 transition where a large fraction of the clusters get interconnected
 into a single cluster which spans across nearly the entire survey
 region \citep{pandey}. Continuing the coarse-graining considerably
 beyond this causes the clusters to thicken and the filamentary
 patterns are washed away. We restrict our analysis to the range $FF
 \le 0.65$ which is slightly beyond the percolation threshold.  A
 point to note here is that the $F_2$ curves are sensitive to the
 area, thickness and number density of the sample \citep{pandey1}
 which varies considerably across the luminosity bins (Table~1), and
 in the present analysis it is not meaningful to compare the $F_2$
 values across the luminosity bins.  Considering any one of the
 luminosity bins in Figure \ref{fig:3}, we find that increasing the
 bias causes $F_2$ to go up at low $FF$ whereas the trend is reversed
 at large $FF$. This can possibly be attributed to the fact that with
 increasing bias the galaxies get concentrated into more compact
 regions at the expense of some of the large-scale coherent
 structures. We also note that the bias dependence of the $F_2$ curves
 is somewhat different from the luminosity dependence \citep{pandey1}
 which shows a monotonic decrease with increasing luminosity.  For
 each luminosity bin we find that the simulated data-set is consistent
 with the actual data for only a small range of the bias parameter.

The results of the quantitative comparison of each luminosity bin
against the biased simulations are shown in Table~2.  It is customary
to use $L/L^*=10^{-0.4(M-M^*)}$ to quantify the luminosity of each
bin, where $M$ is the median magnitude of the bin and $M^*$ has a
value $M^*=-20.44\pm0.01$ \citep{blan3} for the SDSS. The $M^*$
galaxies lie in Bin 3 which has a bias value $b^*=1.1$.  The
luminosity dependence ($L/L^*$ dependence) of the relative bias
$b/b^*$ is shown in Figure \ref{fig:3}.  Fitting a straight line
$b/b^* = A + B \, ( L/L^*)$ by minimizing
\begin{equation}
\chi^2/\nu=\frac{1}{N} \sum^N \frac{[(b/b^*)_{\rm Data}-(b/b^*)_{\rm
    Model}]^2}{\sigma^2_{\rm Data}}
\end{equation} 

we obtain the best fit values $A=0.833 \pm 0.009 $ and $B=0.171 \pm
0.009 $ which gives an acceptable fit with $\chi^2/\nu=0.066$. Here we
have assumed that at each luminosity  bin  $\sigma_{\rm Data}$ is half
the corresponding $95 \%$ confidence interval. Note that such a low
value of $\chi^2/\nu$ is unlikely  (probability $<2.5 \%$) if the
errors in the different luminosity bin are uncorrelated and have a
normal distribution. It is possible that these assumptions do not
hold, and a further possibility is that using half  the 
$95 \%$ confidence interval overestimates $\sigma_{\rm Data}$. 

We find that our best fit
luminosity bias relation is very close to those proposed by
\citet{nor} and \citet{teg2}, both of which provide acceptable fits to
our data with $\chi^2/\nu=0.189$ and $1.154$ respectively. The
luminosity bias relation obtained by \citet{pkok} through fitting the
\citet{beno} data is considerably steeper. Their luminosity bias
relation shows considerable deviations from our data at both the low
and high luminosity ends and is ruled out with
$\chi^2/\nu=4.697$. This is consistent with \citet{nor} who find
considerable deviations between their data and the \citet{beno} data
at high luminosities.

The observed luminosity bias relation is an useful input for models of
galaxy formation.  Hierarchical models of galaxy formation
(e.g. \citealt{white1}) generally assume the brighter galaxies to be
associated with the more massive dark matter halos which have a
stronger clustering. The semi-analytic model of galaxy formation of
\citet{benson} is consistent with the observed luminosity bias
relation \citep{nor}. \citet{zehavi} show that the halo model with
suitably chosen parameters can correctly predict the observed
luminosity dependence. We note that the focus of much of these works
has been on the luminosity dependence of the two-point statistics.
The largest currently available implementation of a semi-analytic
model of galaxy formation (The Millennium Run, \citealt{springel})
fails to correctly predict the luminosity dependence of the
filamentarity for the brightest galaxies \citep{pandey1}. Finally, it
is interesting that our analysis which uses a global property of the
galaxy distribution, namely its morphology, predicts the same
luminosity bias relation as determined from the two point statistics.
  
\section{Acknowledgment}
SB would like to acknowledge financial support from the Govt.  of
India, Department of Science and Technology (SP/S2/K-05/2001). BP
would like to thank the CSIR, Govt. of India for financial support
through a Senior Research fellowship. The SDSS DR4 data was downloaded
from the SDSS skyserver http://skyserver.sdss.org/dr4/en/.

    Funding for the creation and distribution of the SDSS Archive has been 
provided by the Alfred P. Sloan Foundation, the Participating 
Institutions, the National Aeronautics and Space Administration, the 
National Science Foundation, the U.S. Department of Energy, the Japanese 
Monbukagakusho, and the Max Planck Society. The SDSS Web site is 
http://www.sdss.org/.

    The SDSS is managed by the Astrophysical Research Consortium (ARC) for 
the Participating Institutions. The Participating Institutions are The 
University of Chicago, Fermilab, the Institute for Advanced Study, the 
Japan Participation Group, The Johns Hopkins University, the Korean 
Scientist Group, Los Alamos National Laboratory, the Max-Planck-Institute 
for Astronomy (MPIA), the Max-Planck-Institute for Astrophysics (MPA), New 
Mexico State University, University of Pittsburgh, Princeton University, 
the United States Naval Observatory, and the University of Washington.

\end{document}